***The Association Between Commute Time and Poverty in U.S. Counties: Investigating Sprawl***

Ellie Krossa, John Sun, and Topher Wang

May 20, 2018

<u>**Abstract**</u>

Are counties with higher levels of poverty associated with higher average commute times than counties with low levels of poverty and are there other determining factors of commute time? In this research, we attempt to contribute further to research and evidence of the negative effects sprawl causes by assessing county poverty rates affect on commute times. Urban Sprawl is defined as "the expansion of human populations away from central urban centers into low density, car-dependent communities" (Merriam Webster). Thus, our research aims to support and continue research done on sprawl in the past and help aid future research into sprawl that may improve commuters quality of life. This research and its findings could contribute to and promote further investigations of whether people commute further to find work, and if that is caused by a lack of jobs in those counties. If we do find that impoverished counties are associated with higher commute times, there is more evidence that sprawl has taken place. If higher poverty levels do end up being associated with higher county commute times, our research could be extended to a more politically involved agenda of addressing how to reduce high commute time.

<u>**Introduction**</u>

In the last century, there has been a shift in the residential patterns of urban residents. The majority of people lived and worked in central cities. Today, two-thirds of the workers live in the suburbs with many of jobs being located in the suburbs as well. Unfortunately for the other one-third of the workers, they congregate in cities, and cannot move closer to their workplaces that are a concerning distance away from them. Thus, commute times increased for these poor, living in or around cities, due to the changing job market and sprawl (Waller and Sawhill 2018).

A "jobs-housing balance", mentioned by Giuliano and Small (1993), prevents these disadvantaged workers from moving to the expensive suburbs. It supports the argument that the majority of housing around areas of high employment is expensive and inaccessible to the non-wealthy (city-dwellers). This suggests that low income workers will have further commutes to balance commuting cost versus land cost.

David Cotter (2002) states that there are different categories of poverty, "place poverty" and "person poverty". The latter includes risk factors for poverty such as race or gender. The former includes housing and the economic factors that affect communities and may explain mean commute times of workers. Thus, communities with a lack of jobs are more prone to poverty. Cotter's argument contributes to the findings that "poverty rates increase with greater rural distances from metropolitan areas", which in turn leads to difficulty with commuting and the continued impoverishment of non-metropolitan areas with a lack of jobs (Partridge and Rickman 2008).

Sprawl, according to Glaeser and Kahn, is the 21st century phenomenon that some people are not dependent on city-living due to automobiles and therefore can live outside public transportation spheres and cities. This is usually seen as pleasant and accompanied by improved qualities of life, but as they addressed, the problem remains that sprawl causes loss of jobs for those who cannot afford luxurious alternatives but only inferior substitutes (Glaeser and Kahn 2004). Therefore, through our question, we hope to suggest that sprawl has occurred in the U.S. and poverty is one of the consequences.

**Methods:**

The dataset we selected for our research was a collection of United States Census Demographic Data, an observational study. We downloaded the set from Kaggle, but the original data comes from responses to census forms that were distributed in 2015 by the US Census Bureau. The forms were distributed to all households and businesses, and responses to the census are required by law.

Our dataset is aggregate and is a subset of the census data that only contains information for US counties on demographics like racial distribution, income statistics, and work information like commute times and self-employment. We cleaned the data by using the dplyr package to remove some variables we deemed unimportant including raw employment numbers and county names. We also grouped the states into the census regions of West, Midwest, Northeast, South, and other states (Alaska, Hawaii, and Puerto Rico). We then created indicator variables for each of the five regions and added them to the data set.

For missing values of the data, we omitted the counties with NA values in their row. The census data is complete with 3219 observations, one for every U.S. County and 31 variables. We removed  one row/county. We also added new variables to the data set including log of the total population and log of all races except white, log of walk, log of transit to allow us to magnify the effects of the data.

Using the data, we performed single linear regression models to test the preliminary associations and then constructed multiple regression models to fit the model of best fit to answer our original research question.

**Results:**

*Figure 1* shows the normality of the histogram for mean commute time for citizens in each U.S. county, makes it an ideal response variable for constructing a linear model.

*Figure 1.*

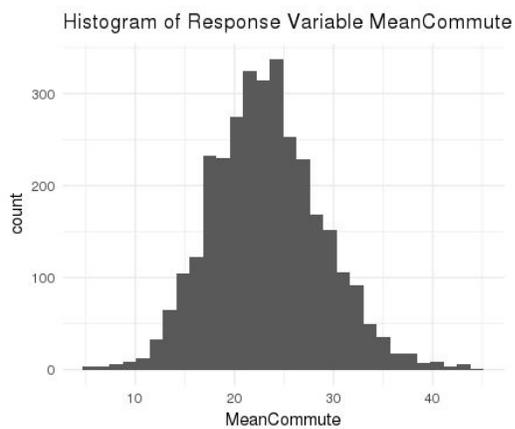

| Mean for County Mean Commute Time | 23.28 |
|---|---|
| Standard Deviation for County Mean Commute Time | 5.6 |

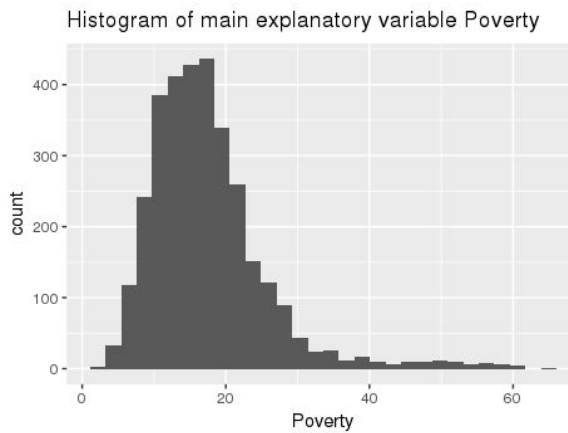

*Figure 2.*

| Median Poverty Value | 16.2 |
|---|---|
| IQR for Poverty variable | 8.6 |

Given the skewed histogram of our primary explanatory variable (*Figure 2*), we chose to report median (center) and IQR (spread) below Figure 2 instead of mean and standard deviation, statistics that are less resistant to extreme observations.

*Table 1.*

| Variable | Description | Median | Mean | SD |
|---|---|---|---|---|
| *Asian* | % County Asian | .5 | 1.223 | 2.609 |
| *Black* | % County Black | 1.9 | 8.668 | 14.28 |
| *Carpool* | % People that Carpool to work | 9.9 | 10.27 | 2.907 |
| *Childpoverty* | % Children in Poverty | 22.7 | 24.18 | 11.69 |
| *Construction* | % of People working in Construction | 12.1 | 12.71 | 4.215 |
| *Drive* | % People that Drive to work | 80.7 | 79.19 | 7.618 |
| *Hispanic* | % County Hispanic | 3.9 | 11.01 | 19.24 |
| *IncomePerCap* | Income Per Capita in County | 23459 | 23970 | 6192 |
| *lnTotalPop* | Log of Total Population | 10.168 | 10.27 | 1.460 |
| *MeanCommute* | Mean Commute Time in minutes | 23 | 23.28 | 5.596 |
| *Men* | Number of Men in County Pop. | 12944 | 48910 | 15670 |
| *Native* | % County Native | .3 | 1.724 | 7.254 |
| *Office* | % People working in office work | 22.4 | 22.22 | 3.2 |
| *Pacific* | % County Pacific Islander | 0 | .0717 | .3934 |
| *Poverty* | % County in Poverty | 16.2 | 17.49 | 8.319 |
| *Production* | % County Working in Production | 15.2 | 15.73 | 5.737 |
| *Professional* | % People in Professional jobs | 29.9 | 30.99 | 6.369 |
| *Service* | % people working in Service jobs | 18.1 | 18.34 | 3.635 |
| *TotalPop* | Total Population | 26056 | 9944 | 31940 |
| *Transit* | % People taking Transit to work | .4 | .9721 | 3.059 |
| *Unemployment* | % People Unemployed | 7.6 | 8.097 | 4.049 |
| *Walk* | $ People who walk to work | 2.4 | 3.312 | 3.699 |
| *White* | % County White | 84.1 | 75.44 | 22.93 |
| *Women* | Number of women in County | 13063 | 50520 | 16270 |

In *Table 1,* we created a table of summary statistics for the 22 quantitative variables in our dataset.

<u>*Figure 3.*</u>

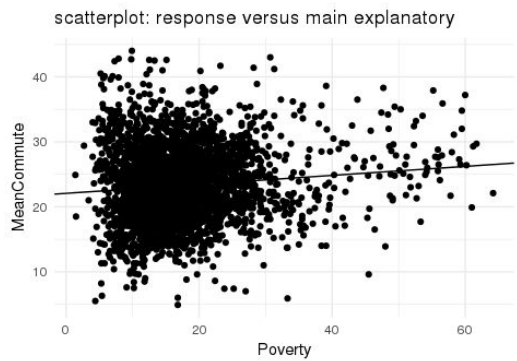

*Figure 3* produced an r² of 0.01 between MeanCommute and Poverty. The sum of squares explained is 1% of the total variability in MeanCommute. This shows a feeble association between commute time and poverty. This was fairly low when compared to the $r^2$ of something like unemployment that explained more variability at 8.6%.

<u>*Table 2.*</u>

| SLR Predictor Variable | R-squared | F-statistic | p-value |
|---|---|---|---|
| Poverty | 0.001 | 34.3 | 5.21e-09 |
| Professional | 0.009 | 29.9 | 4.99e-08 |
| Office | 0.036 | 121.5 | 2.20e-16 |
| Drive | 0.052 | 177.3 | 2.20e-16 |
| Unemployment | .086 | 301.3 | 2.20e-16 |
| logTotalPop | 0.073 | 254.1 | 2.20e-16 |
| logHispanic | 0.002 | 5.8 | 1.65e-02 |
| logBlack | 0.061 | 208.4 | 2.20e-16 |
| logNative | 0.076 | 264.3 | 2.20e-16 |
| logPacific | 0.009 | 29.2 | 7.19e-08 |
| White | .006 | 19.9 | 8.36e-08 |

In *Table 2,* we ran simple linear regressions looking at the relationships between our variables and the mean commute time of counties. The SLR predicted MeanCommute using each element of the set of predictor variables that we expected to be associated with commute time. The results are shown above in Table 2.

*Figure 4. Distribution of Mean Commute Time(minutes) for all 5 regions*

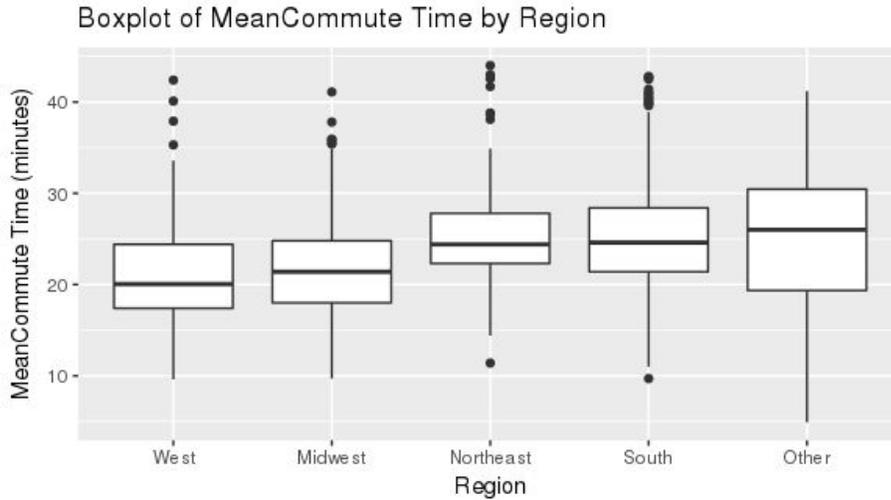

After performing an ANOVA we found an F-statistic of 93.17 and a p-value of 2e-16. We conclude that at least two of the mean commutes among the 5 regions are significantly different from each other.

*Table 3: Summary of MLR*

**Mean Commute** = lnNative + lnTransit + lnPacific + lnTotalPop + lnAsian + lnWalk + lnBlack + White + lnHispanic + i.West + i.Midwest + i.Northeast + i.otherRegion + Uemployment + Carpool + Production + Construction + Office + Service + Professional + ChildPoverty + Poverty + IncomeperCap + Drive

*\*Baseline=i.South*

|  | MeanCommute | | |
|---|---|---|---|
|  | Estimate | Conf. Int. | p-value |
| (Intercept) | -72.65 | -301.44 – 156.14 | .534 |
| lnNative | -1.57 | -1.81 – -1.32 | <.001 |
| lnTransit | 2.45 | 2.02 – 2.87 | <.001 |
| lnPacific | -0.99 | -2.10 – 0.12 | .079 |
| lnTotalPop | 0.67 | 0.48 – 0.85 | <.001 |
| lnAsian | -1.26 | -1.71 – -0.81 | <.001 |
| lnWalk | -4.57 | -5.03 – -4.11 | <.001 |
| lnBlack | -0.48 | -0.70 – -0.26 | <.001 |
| White | -0.00 | -0.02 – 0.02 | .961 |
| lnHispanic | -1.21 | -1.46 – -0.95 | <.001 |
| i.west | -1.62 | -2.25 – -1.00 | <.001 |
| i.midwest | -2.06 | -2.52 – -1.60 | <.001 |
| i.northeast | 0.27 | -0.48 – 1.02 | .481 |
| i.otherRegion | 2.12 | 0.85 – 3.39 | .001 |
| Unemployment | 0.39 | 0.33 – 0.45 | <.001 |
| Carpool | 0.06 | -0.00 – 0.13 | .058 |
| Production | 0.97 | -1.32 – 3.26 | .408 |
| Construction | 1.20 | -1.09 – 3.49 | .305 |
| Office | 0.99 | -1.30 – 3.28 | .395 |
| Service | 0.97 | -1.32 – 3.26 | .406 |
| Professional | 0.98 | -1.31 – 3.27 | .400 |
| ChildPoverty | -0.02 | -0.06 – 0.02 | .259 |
| Poverty | -0.01 | -0.08 – 0.05 | .668 |
| IncomePerCap | 0.00 | 0.00 – 0.00 | .011 |
| Drive | -0.08 | -0.12 – -0.04 | <.001 |
| Observations | 3219 | | |
| $R^2$ / adj. $R^2$ | .438 / .434 | | |

Summary of MLR in *Table 3*: The coefficients of the variables: Production, Construction, Office, Service, Professional, ChildPoverty, Poverty are not significantly different from zero. The model explains 43.8 percent of the total variability in MeanCommute. Betas of all other variables are significantly different from zero. Unemployment and income per capita are significant indicators though.

*Table 4: Summary of Final MLR Model*

| Term | Coefficient | Std Err | t-value | p-value |
|---|---|---|---|---|
| (Intercept) | 31.66014 | 1.82777 | 17.322 | < 2e-16 *** |
| lnNative | -1.24992 | 0.11981 | -10.433 | < 2e-16 *** |
| lnTransit | 1.46206 | 0.58197 | 2.512 | 0.012045 * |
| lnPacific | -1.11888 | 0.54392 | -2.057 | 0.039761 * |
| lnTotalPop | 0.70983 | 0.08677 | 8.181 | 4.03e-16 *** |
| lnAsian | -1.16539 | 0.42008 | -2.774 | 0.005566 ** |
| lnWalk | -1.75621 | 0.56256 | -3.122 | 0.001813 ** |
| lnBlack | -0.33905 | 0.08996 | -3.769 | 0.000167 *** |
| lnHispanic | -2.01440 | 0.15620 | -12.896 | < 2e-16 *** |
| i.westTRUE | -1.55090 | 0.29338 | -5.286 | 1.33e-07 *** |
| i.midwestTRUE | -2.41798 | 0.20207 | -11.966 | < 2e-16 *** |
| i.otherRegionTRUE | 1.62189 | 0.66248 | 2.448 | 0.014410 * |
| Unemployment | 0.32060 | 0.06489 | 4.941 | 8.18e-07 *** |
| Construction | 0.22350 | 0.02365 | 9.451 | < 2e-16 *** |
| Poverty | -0.09505 | 0.01453 | -6.543 | 7.01e-11 *** |
| Drive | -0.12864 | 0.01849 | -6.958 | 4.17e-12 *** |
| I(lnWalk^2) | -0.80928 | 0.18616 | -4.347 | 1.42e-05 *** |
| lnWalk:Unemployment | -0.04403 | 0.02993 | -1.471 | 0.141405 |
| lnTransit:Unemployment | 0.06284 | 0.03943 | 1.594 | 0.111041 |
| lnTransit:lnWalk | -1.11004 | 0.25390 | -4.372 | 1.27e-05 *** |
| lnHispanic:Unemployment | 0.08089 | 0.01456 | 5.556 | 2.99e-08 *** |
| lnTransit:lnAsian | 1.78222 | 0.17711 | 10.063 | < 2e-16 *** |
| lnAsian:Unemployment | -0.15954 | 0.04488 | -3.555 | 0.000383 *** |
| | | | | |
| Multiple R-Sq | 0.4688 | | Adj R-Sq | 0.4652 |

The final model in *Table 4* shows the model with the best possible fit using second order terms. It shows that our explanatory variable, Poverty, is highly associated with Mean Commute time, and several other variables including unemployment are shown to be significant predictors as well. The model in total explains 46.88% of the total variability.

*Table 5: Comparison of Models*

| Model Name | A.I.C. | R-squared | Adjusted R-squared | $\mu$ of X-validation Error |
|---|---|---|---|---|
| Model 1: Initial | 18415 | 0.4384 | 0.4342 | 17.95708 |
| Model 2: Reduced | 18405 | 0.4377 | 0.4347 | 17.89115 |
| Model 3: Full | 18234 | 0.4690 | 0.4650 | 16.97825 |
| Model 4: Final | 18232 | 0.4688 | 0.4651 | 16.95901 |

Our first model included all the variables in our dataset. We used stepwise elimination on the first model to produce model 2. We then made a regression tree using the reduced model and added all the interaction terms from the regression tree to make model 3. The regression tree adds term it thinks would minimize cross validation error. Finally, we applied stepwise elimination to model 3 to produce model 4. Stepwise elimination drops each term in the model and checks which term(s) to eliminate to minimize the A.I.C.

**Nested ANOVA tests:**

Conclusions of Nested ANOVA comparisons and rankings of models are an equivalence relation that is reflexive, symmetric and transitive. According to ANOVA, Model $1 \leq$ Model2 $\leq$ Model3 $\leq$ Final Model. The p-values from nested ANOVAs (in order) are .7866, 0, and 0.514. The F-statistics in order are 0.5628, 26.88, and 0.6653. The first p-value indicates that Model One is as good better than Model 2. The second p-value indicates that Model 3 is as good or better than Model 2. The third p-value indicates that the final model as good or better than Model 3. AIC and X-validation error decreasing while Multiple-R-squared is increasing implies that we made a better model in each step.

**<u>Discussion:</u>**

Based on our final model, we found that poverty is associated with mean commute time for U.S. counties. We found that as a county's poverty rate increases, the county's average commute time will decrease. This association seems to be highly significant with a p-value of 7.01e-11 after controlling for other variables. This association is actually the opposite of what we expected. We expected that as poverty rates increase in counties, their mean commutes would also increase. Yet, the results indicate the opposite association, where mean commute tends to go down.

On the other hand, we did find that there was a highly significant association between unemployment and mean commute time. The model showed that as unemployment levels rose, a county's mean commute time would also rise. This indicates that in counties with high levels of unemployment, people are traveling greater amounts of time to get to work.

As we analyze our findings, we speculate as to why our initial hypothesis is opposite of the resulting effect showing a negative association between poverty and mean commute. One factor we did not consider is that almost 40% of people living in poverty are not currently working, which means that in counties with high levels of poverty, the average commute time would be negatively affected (Gould 2015). Another explanation for our unexpected result may be that mean commute time does not account for traffic. In high density areas, traffic may drastically increase commute time, even if the actual distance to work is not very far.

As for support that spawl is occurring, the unemployment variable discussed earlier indicates that areas with higher levels of unemployment also have higher commute times. Thus, upon further investigation we could possibly find that these areas with high commute times have less employment opportunities, which is why people are commuting further. As Waller and Sawhill (2018) argued, as sprawl takes place, the poor are left having to travel great distances to find work if they are even able to.

Majeski (2016) tested how individual commute times were associated with the likelihood that an individual lived in poverty. Contradictory to our results, he found that there was a significant increase in poverty levels as commute time increased. This could be attributed to the fact that he is looking at individual data compared to our study looking at aggregate county data.

It is also important to note that the results from our final model are different than our results from our initial model. Originally, in our linear regression model, we found a slightly positive relationship between poverty and commute time. But, it was not significant. And then again, when we produced our initial multiple regression model, we found that there was no significant association after controlling for the other variables. However, when we included interaction and second order terms, poverty became highly significant in predicting mean commute time.

Our results regarding poverty seem to conflict with research on sprawl that argue jobs are moving away from the poor and average commute times are increasing. But, our results for unemployment confirm the existence of sprawl and implicate that the poor are living in areas that do not have as many jobs, thus more people are unemployed. This argument aligns with previous research that states that unemployment is a verified consequence of sprawl, which affects the poor that are left behind (Bhatta 2010).

For further studies, conducting individual level analysis on poverty and mean commute time may help to solidify a stronger association between the two. Other important considerations would be to include data on number of available jobs in the county, whether the county is a metropolitan area or a rural area, and the ability to restrict the data to people that are employed, rather than having the data for unemployed people affecting commute time, like we believe it did in ours.

This research has brought up questions such as: does the region of the county affect its commute times, does the type of area affect them, and do traffic levels affect them? With all of these questions, the argument for sprawl's effect on counties could be expanded and new considerations could be addressed. In total, although the results are opposite of what we would have expected, there does seem to be strong evidence that suggests sprawl has occurred in counties across the U.S. using unemployment and poverty rates.

## Bibliography

Cotter, David A. (2002). "Poor People in Poor Places: Local Opportunity Structures and
Household Poverty." *Rural Sociology,* Vol 67, No. 4: 534-555.

Bhatta, B. (2010). *Causes and Consequences of Urban Growth and Sprawl*. Berlin:
Springer-Verlag.

Bolton, Roger (1992). "Place Prosperity vs. People Prosperity Revisited: An Old Issue with a
New Angle". *Urban Studies*, Vol 29, No. 2: 185-203.

Giuliano, Genevieve & Small, Kenneth (1993). "Is the Journey to Work Explained by Urban
Structure?" *Urban Studies,* Vol 30, No. 9: 1485-1500.

Glaeser, Edward L. & Kahn, Matthew E. (2004). "Sprawl and Urban Growth." *Handbook of
Regional and Urban Economics,* Vol. 4: 2481-2527.

Gobillion, Laurent & Selod, Harris & Zenou, Yves (2007). "The Mechanisms of Spatial
Mismatch." *Urban Studies,* Vol. 44, No. 12: 2401-2427.

Gould, Elise (2015). "Poor People Work." *Economic Policy Institute.* May 19, 2015.

Majeski, Quinn (2016). "Effect of Commute Time on Poverty in the United States." *Evans
School of Public Policy Review*, Volume 6.

Partridge, Mark D. & Rickman, Dan S. (2008). "Distance from Urban Agglomeration Economies
and Rural Poverty." *Journal of Regional Science,* Vol. 48, No. 2: 285-310.

Waller, Margy & Sawhill, Isabel (2018). "High Cost or High Opportunity Cost? Transportation
and Family Economic Success." Center on Families and Children, The Brookings
Institution.

Ellie Krossa, Topher Wang, John Sun

April 20, 2018

SC321 Project Part 3

**Appendix:**

**Step 1:** We start by examining the distributions of our categorical variables. Figure 1 shows the bar chart for the indicator variables for region. Table 1 shows the number of observations in each category. Figure 2 shows the histograms of the quantitative variables.

***Categorical Variables:*** *Figure 1*

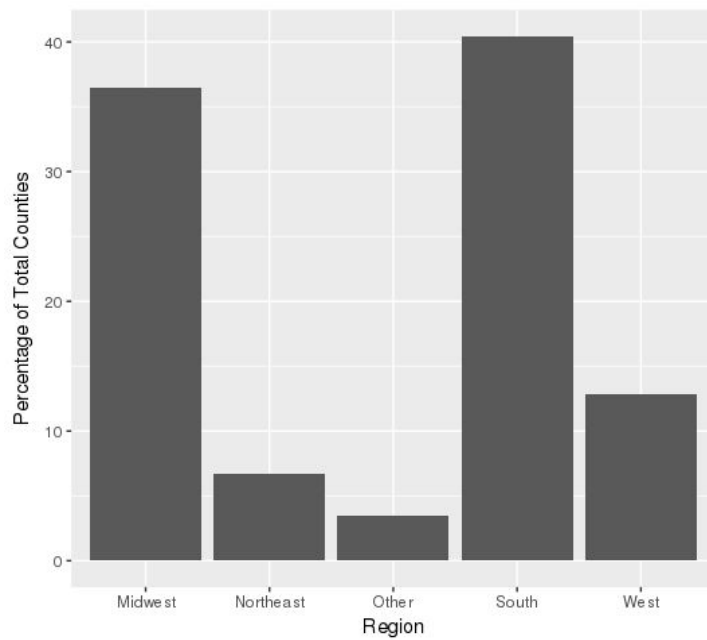

South is the most frequent region and baseline.

*Table 1*

| | Region | NumCounties | PercentCounties |
|---|---|---|---|
| 1 | Midwest | 1175 | 36.502019 |
| 2 | Northeast | 217 | 6.741224 |
| 3 | South | 1302 | 40.447344 |
| 4 | West | 414 | 12.861137 |
| 5 | Other | 111 | 3.448276 |

***Quantitative Variables:*** *Figure 2*

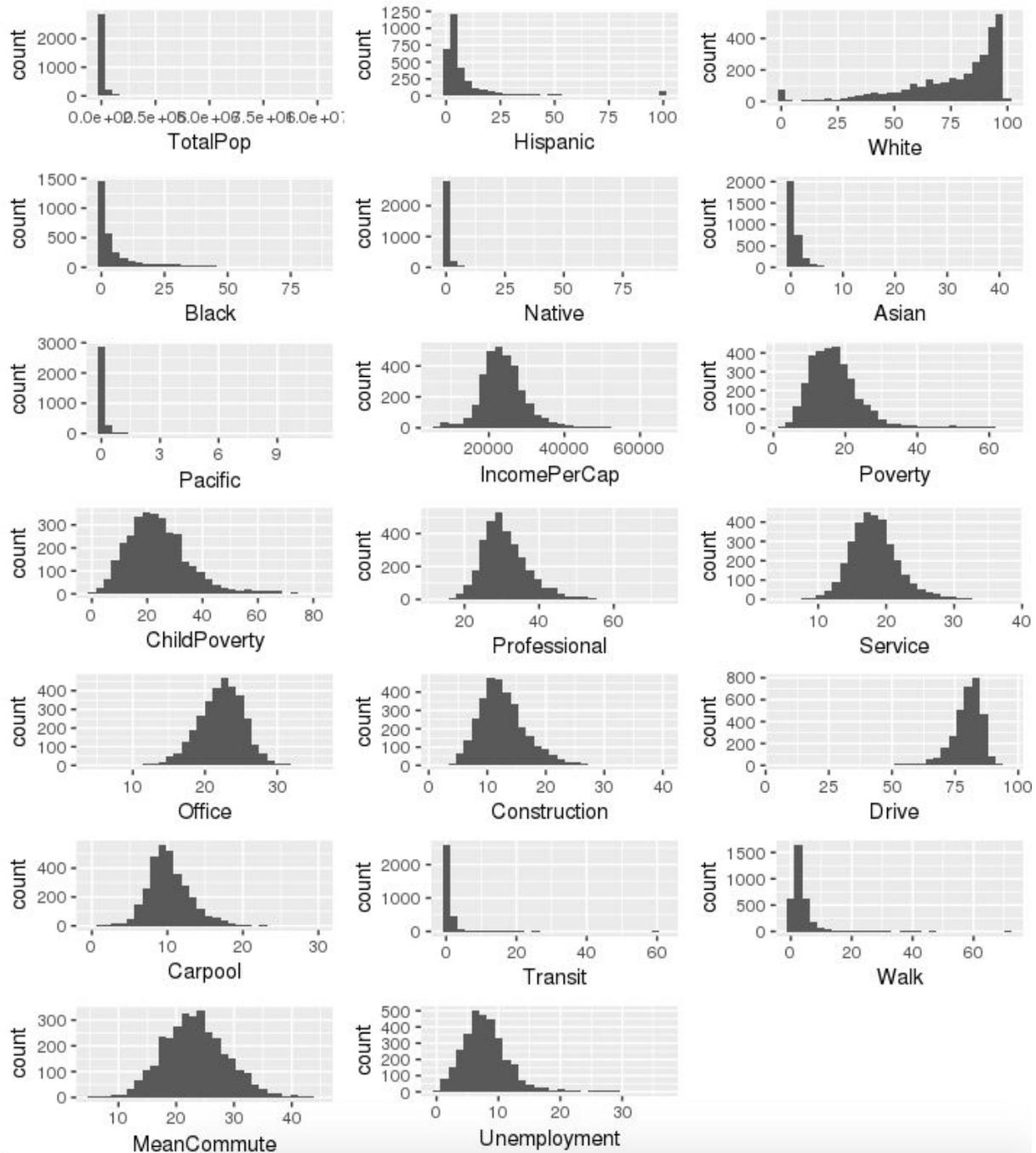

These variables include types of jobs people have in the county, how they get to work, unemployment levels, the percentage of each race in the county, as well as variables for poverty and unemployment.

For the variables that were not normally distributed except White, we used a log transformations. These variables are the race variables of Black, Hispanic, Asian, and Pacific, Native, and the variables of walk, Transit, and Total Population. Transit and Native improved after log transform but remained skewed. Drive, Pacific, Asian, Black also remained skewed as their population is low.

***Transformed Variables:*** *Figure 3*

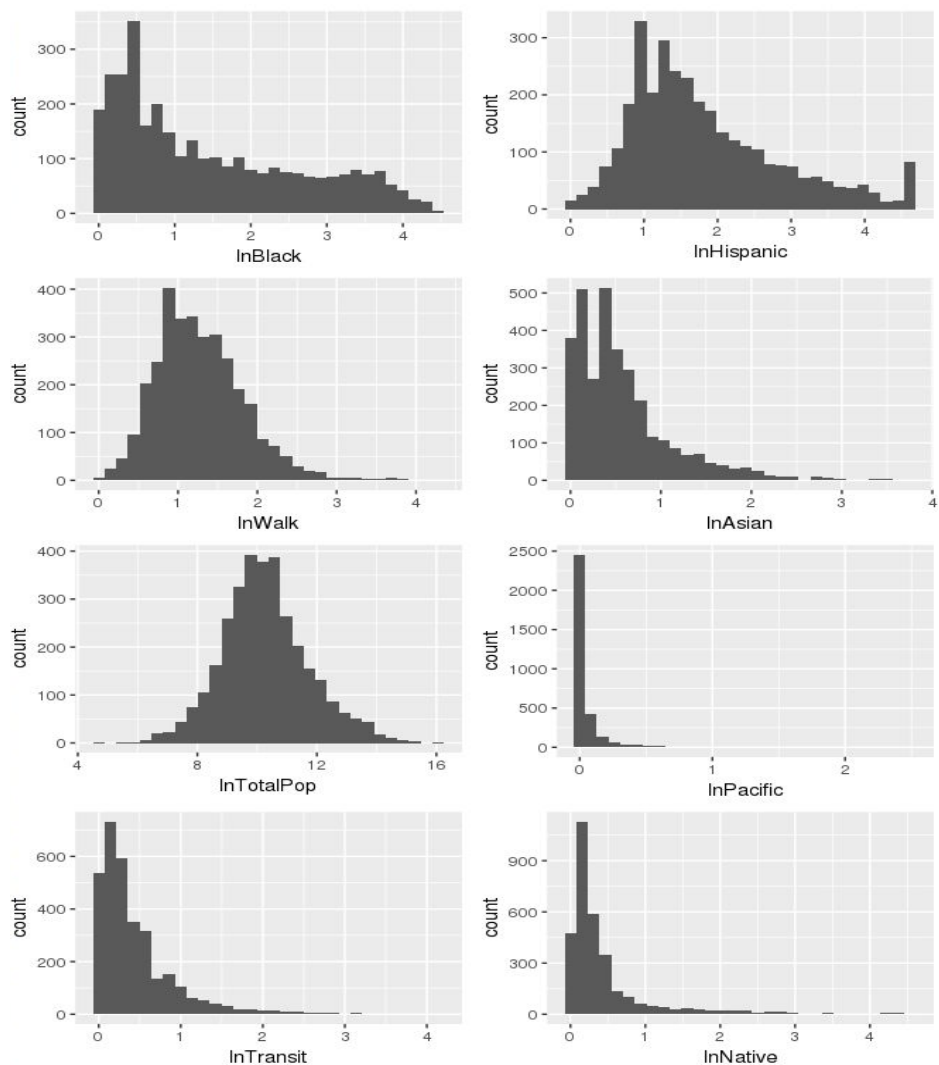

**Step 2:** Next, Table 2 provides the summary statistics for all the quantitative variables in our data.

***Summary Statistics:*** *Table 2*

| Variable | | Median | Mean | SD |
|---|---|---|---|---|
| *Asian* | % County Asian | .5 | 1.223 | 2.609 |
| *Black* | % County Black | 1.9 | 8.668 | 14.28 |
| *Carpool* | % People that Carpool to work | 9.9 | 10.27 | 2.907 |
| *Childpoverty* | % Children in Poverty | 22.7 | 24.18 | 11.69 |
| *Construction* | % of people working in Construction | 12.1 | 12.71 | 4.215 |
| *Drive* | % People that Drive to work | 80.7 | 79.19 | 7.618 |
| *Hispanic* | % County Hispanic | 3.9 | 11.01 | 19.24 |
| *IncomePerCap* | Income Per Capita in County | 23459 | 23970 | 6192 |
| *lnTotalPop* | Log of Total Population | 10.168 | 10.27 | 1.460 |
| *MeanCommute* | Mean Commute Time in minutes | 23 | 23.28 | 5.596 |
| *Men* | Number of Men in County Pop. | 12944 | 48910 | 15670 |
| *Native* | % County Native | .3 | 1.724 | 7.254 |
| *Office* | % People working in office work | 22.4 | 22.22 | 3.2 |
| *Pacific* | % County Pacific Islander | 0 | .0717 | .3934 |
| *Poverty* | % County in Poverty | 16.2 | 17.49 | 8.319 |
| *Production* | % County Working in Production | 15.2 | 15.73 | 5.737 |
| *Professional* | % People in Professional jobs | 29.9 | 30.99 | 6.369 |
| *Service* | % People Working in Service jobs | 18.1 | 18.34 | 3.635 |
| *TotalPop* | Total Population | 26056 | 9944 | 31940 |
| *Transit* | % People Taking Transit to Work | .4 | .9721 | 3.059 |
| *Unemployment* | % People Unemployed | 7.6 | 8.097 | 4.049 |
| *Walk* | $ People who walk to work | 2.4 | 3.312 | 3.699 |
| *White* | % County White | 84.1 | 75.44 | 22.93 |
| *Women* | Number of women in County | 13063 | 50520 | 16270 |

**Step 3**: The scatterplots below (Figure 4) examine the associations between Mean Commute time, our response variable, and the quantitative explanatory variables. Then we produce an analysis of variance test for the simple linear regression model to see if each variable is useful.

***Scatterplots****: Figure 4*

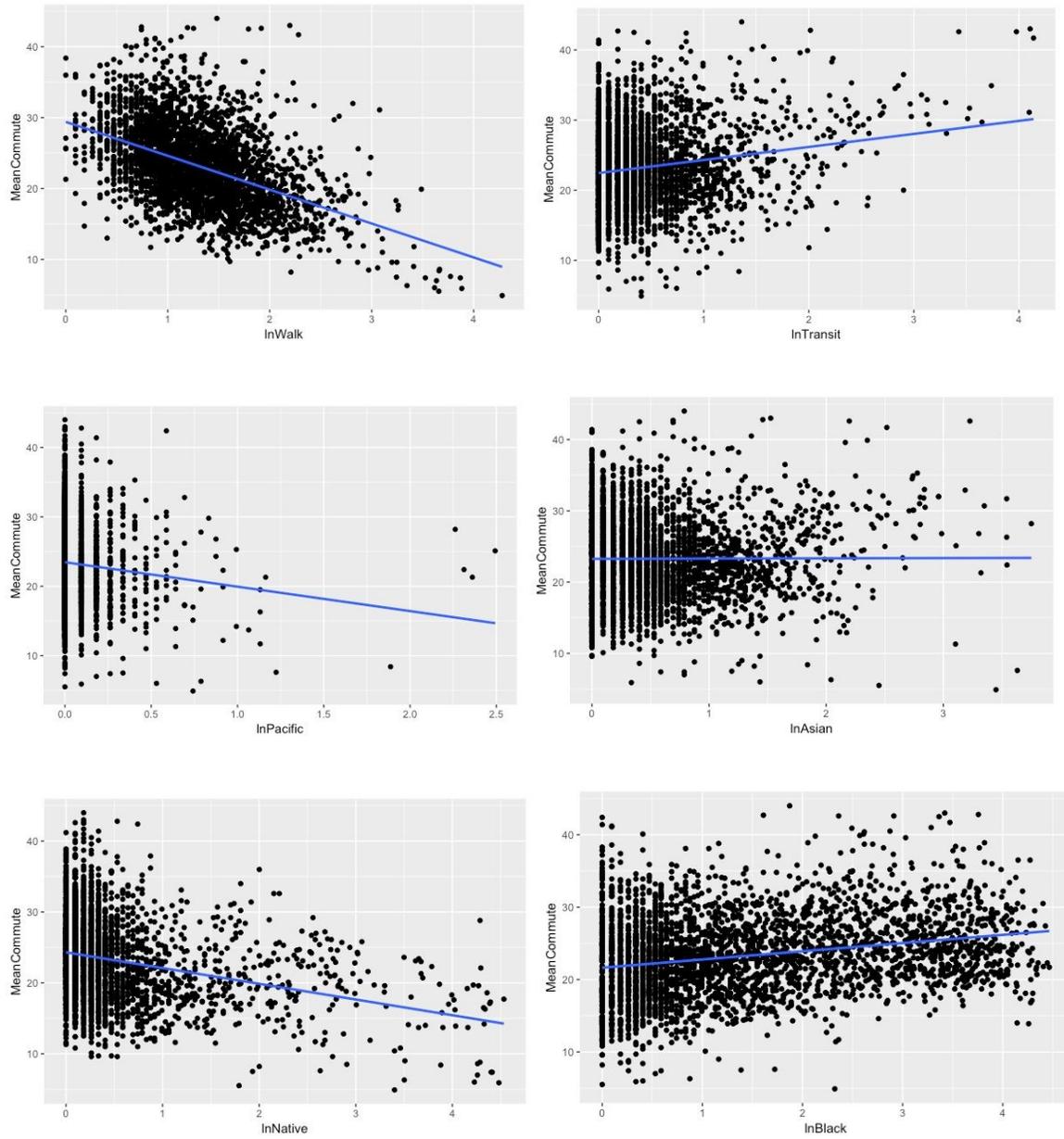

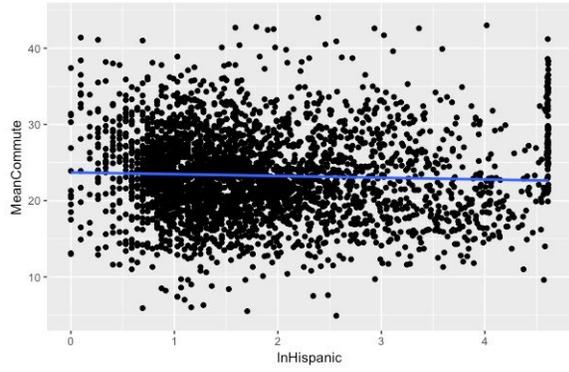
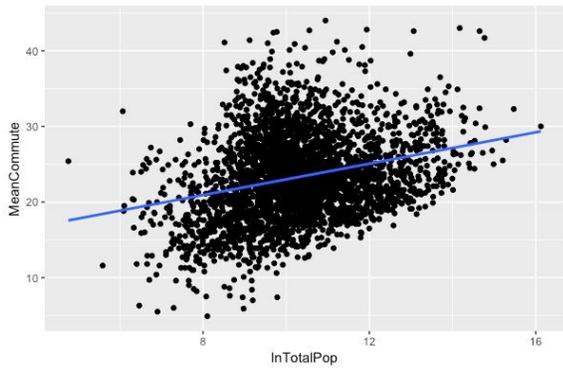
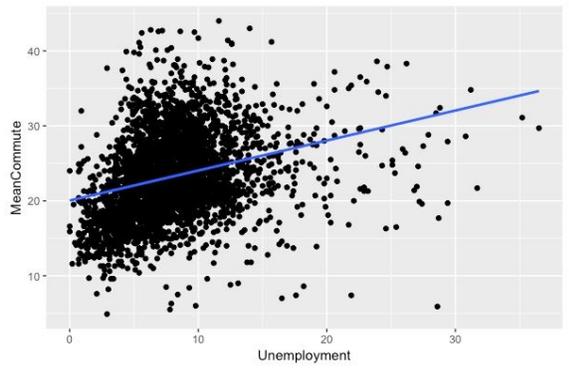
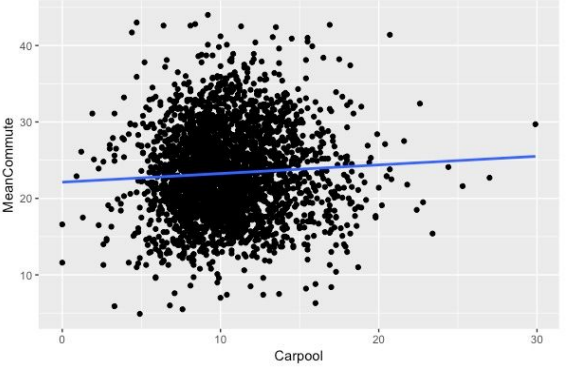
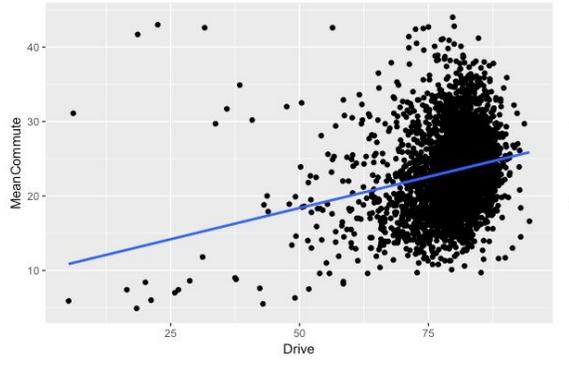
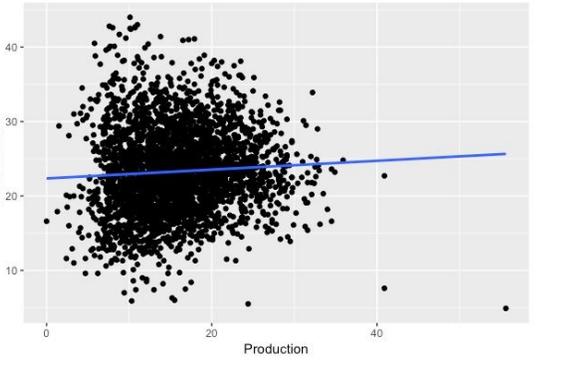
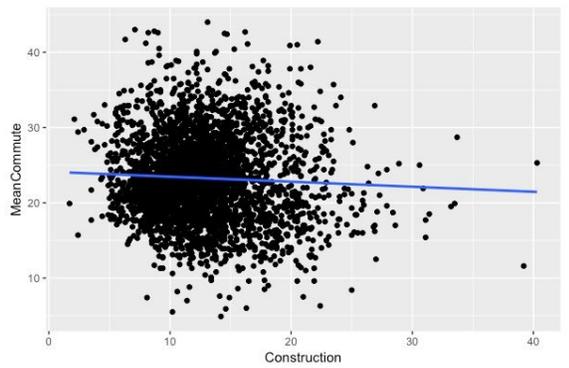
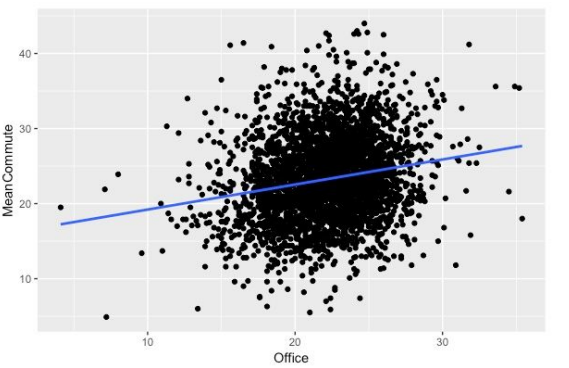

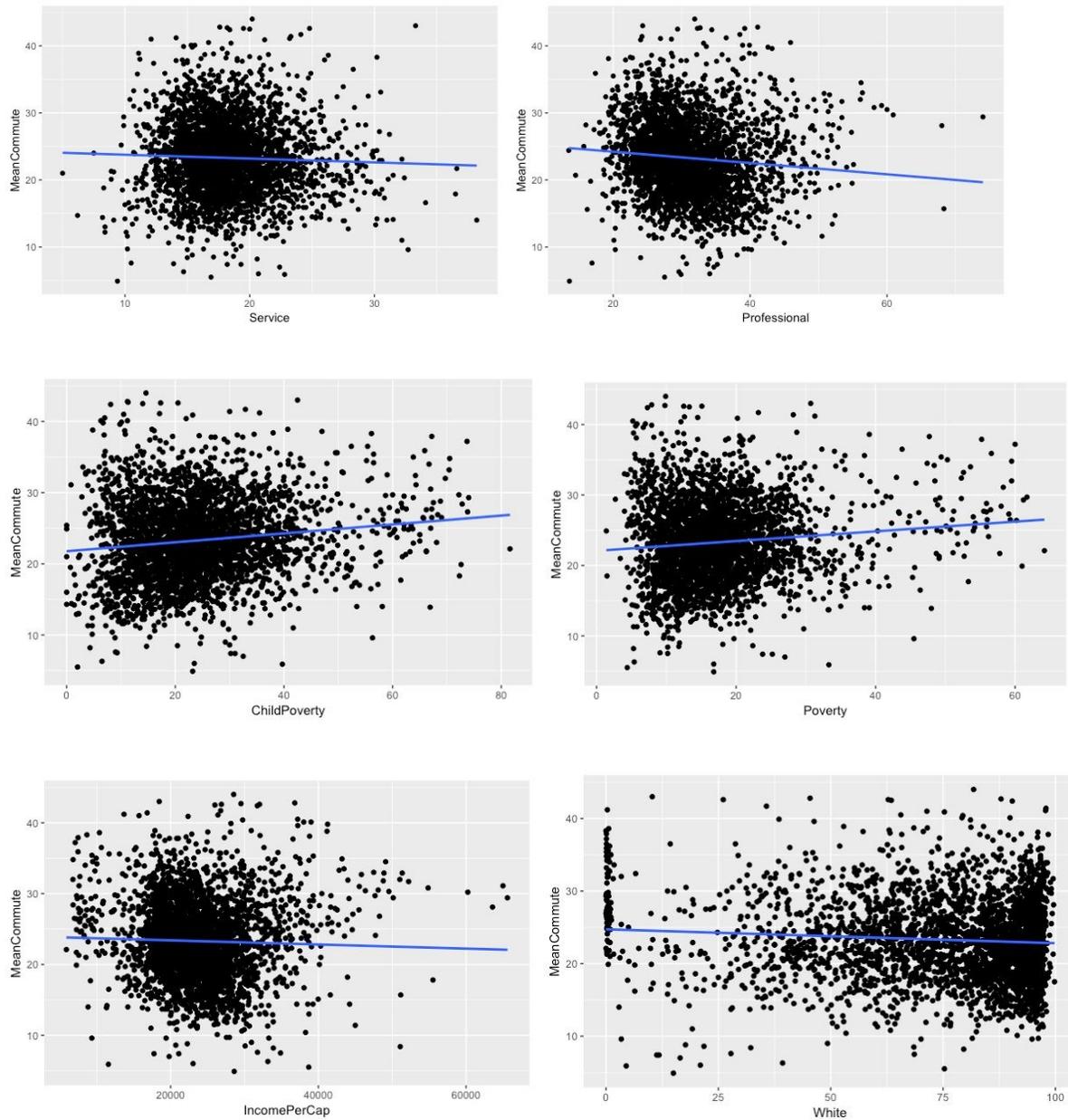

*Table 3* below summarizes the results of the statistical tests for simple linear model using our explanatory variables to predict MeanCommute.

*Table 3*

| SLR Predictor Variable | R-squared | F-statistic | p-value |
|---|---|---|---|
| Poverty | 0.001 | 34.3 | 5.21e-09 |

| Professional | 0.009 | 29.9 | 4.99e-08 |
|---|---|---|---|
| Office | 0.036 | 121.5 | 2.20e-16 |
| Drive | 0.052 | 177.3 | 2.20e-16 |
| Unemployment | .086 | 301.3 | 2.20e-16 |
| logTotalPop | 0.073 | 254.1 | 2.20e-16 |
| logHispanic | 0.002 | 5.8 | 1.65e-02 |
| logBlack | 0.061 | 208.4 | 2.20e-16 |
| logNative | 0.076 | 264.3 | 2.20e-16 |
| logPacific | 0.009 | 29.2 | 7.19e-08 |
| White | .006 | 19.9 | 8.36e-08 |

Then, in Figure 5 and Table 4, we produced the boxplots for our categorical variables and performed an ANOVA (for difference in means).

*Figure 5*

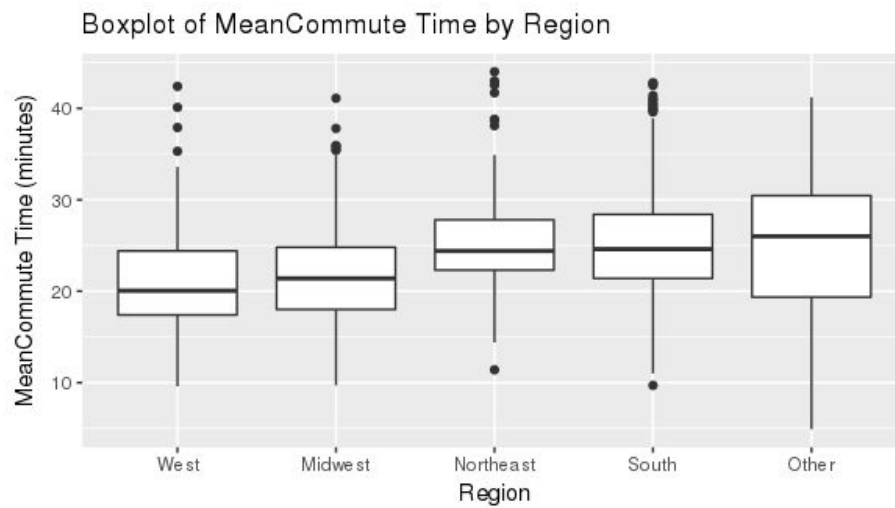

*ANOVA Results:*

There is a significant difference in mean commute time between at least two of the 5 groups, F=93.17, p<2E-16. We shall make a model using south as the baseline. We expect our model to drop i.Northeast as i.Northeast seems no different from the baseline.

**Step 4**: We produced a correlation matrix in Figure 6 and a scatterplot matrix in Figure 7 below to examine whether there is any correlation between the quantitative explanatory variables. The collinear variables include the most shaded.

***Correlation Matrix***: *Figure 6*

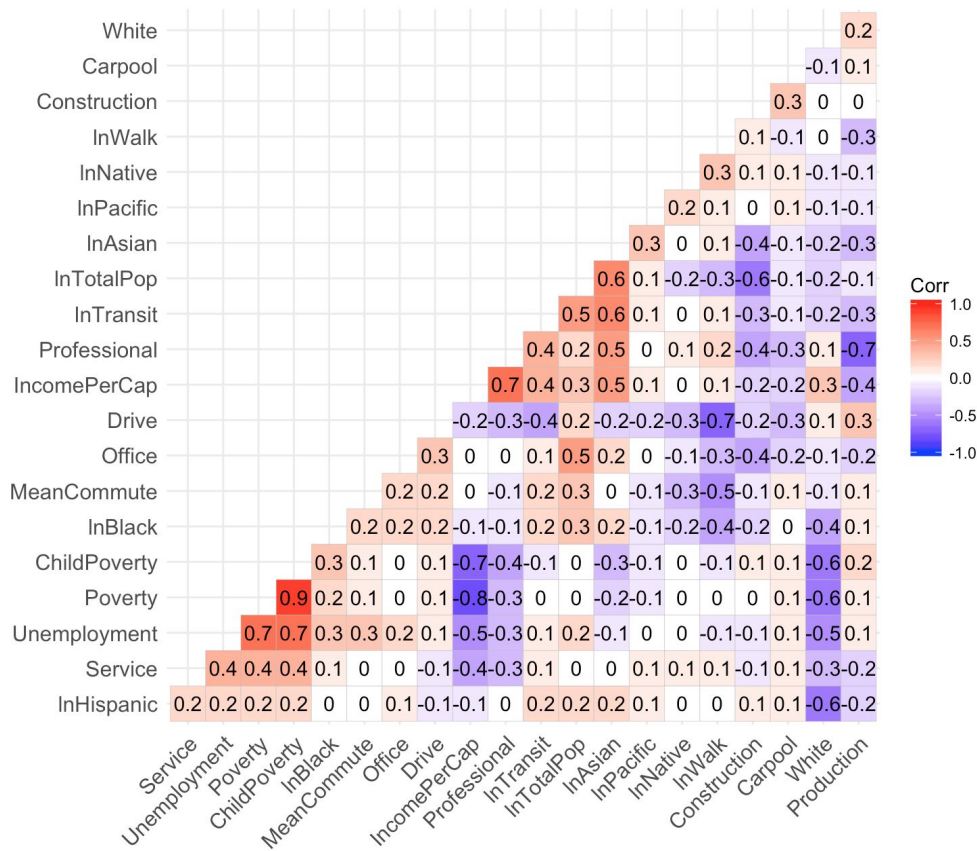

The legend's color saturation tells us the strength of correlation.

***Scatterplot Matrix****: Figure 7*

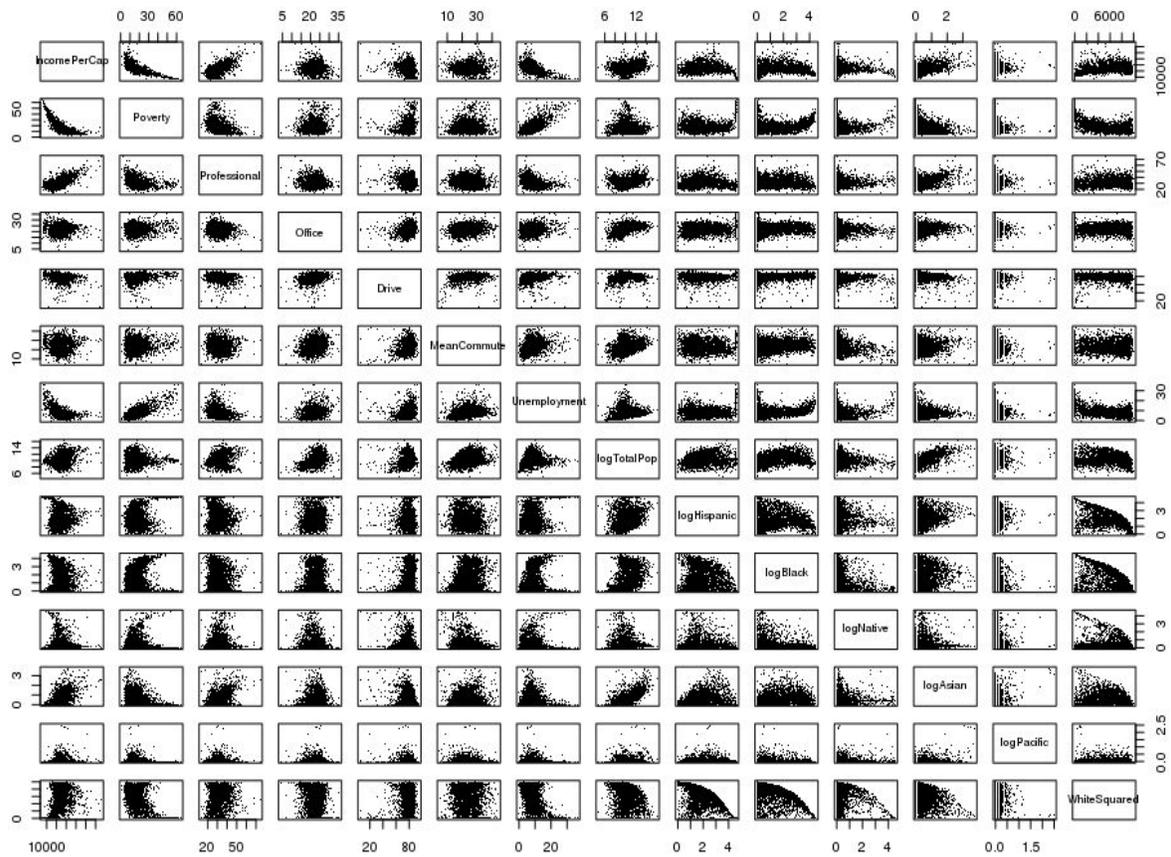

**Step 5:** Using our variables, we produced an initial multiple regression model in Figure 8 and the

summary of the model in Figure 9. We also included the diagnostic plots of our MLR in Figure

10.

*Figure 8*

Let i.West+i.Midwest+i.Northeast+i.otherRegion+i.South=1.

**Mean Commute =** lnNative + lnTransit + lnPacific + lnTotalPop + lnAsian + lnWalk + lnBlack
+ White + lnHispanic + i.West + i.Midwest + i.Northeast + i.otherRegion + Uemployment +
Carpool + Production + Construction + Office + Service + Professional + ChildPoverty +
Poverty + IncomeperCap + Drive.

**MLR Summary**: *Figure 9*

| | Estimate | Conf. Int. | p-value |
|---|---|---|---|
| | | *MeanCommute* | |
| (Intercept) | -72.65 | -301.44 — 156.14 | .534 |
| lnNative | -1.57 | -1.81 — -1.32 | <.001 |
| lnTransit | 2.45 | 2.02 — 2.87 | <.001 |
| lnPacific | -0.99 | -2.10 — 0.12 | .079 |
| lnTotalPop | 0.67 | 0.48 — 0.85 | <.001 |
| lnAsian | -1.26 | -1.71 — -0.81 | <.001 |
| lnWalk | -4.57 | -5.03 — -4.11 | <.001 |
| lnBlack | -0.48 | -0.70 — -0.26 | <.001 |
| White | -0.00 | -0.02 — 0.02 | .961 |
| lnHispanic | -1.21 | -1.46 — -0.95 | <.001 |
| i.west | -1.62 | -2.25 — -1.00 | <.001 |
| i.midwest | -2.06 | -2.52 — -1.60 | <.001 |
| i.northeast | 0.27 | -0.48 — 1.02 | .481 |
| i.otherRegion | 2.12 | 0.85 — 3.39 | .001 |
| Unemployment | 0.39 | 0.33 — 0.45 | <.001 |
| Carpool | 0.06 | -0.00 — 0.13 | .058 |
| Production | 0.97 | -1.32 — 3.26 | .408 |
| Construction | 1.20 | -1.09 — 3.49 | .305 |
| Office | 0.99 | -1.30 — 3.28 | .395 |
| Service | 0.97 | -1.32 — 3.26 | .406 |
| Professional | 0.98 | -1.31 — 3.27 | .400 |
| ChildPoverty | -0.02 | -0.06 — 0.02 | .259 |
| Poverty | -0.01 | -0.08 — 0.05 | .668 |
| IncomePerCap | 0.00 | 0.00 — 0.00 | .011 |
| Drive | -0.08 | -0.12 — -0.04 | <.001 |
| Observations | | 3219 | |
| $R^2$ / adj. $R^2$ | | .438 / .434 | |

i.south is the baseline variable. The initial model has 24 variables. higher Poverty is associated with decreased commute time. The number of whites is not associated with commute time. Asian is associated with lower commute time. Midwest is associated with the lowest commute time compared to southerners. Construction workers are associated with higher commute time. Income per cap is not associated with commute time.

### Diagnostic Plots for Initial MLR

We assume linearity and we have a linear combination of variables for the formula for MLR. We assume the independence of observations. The residuals are normally distributed. Given the constant band in the residuals versus fitted plot, the equal variance assumption is met. We assume simple random sample. Refer to *Figure 10*.

*Figure 10*

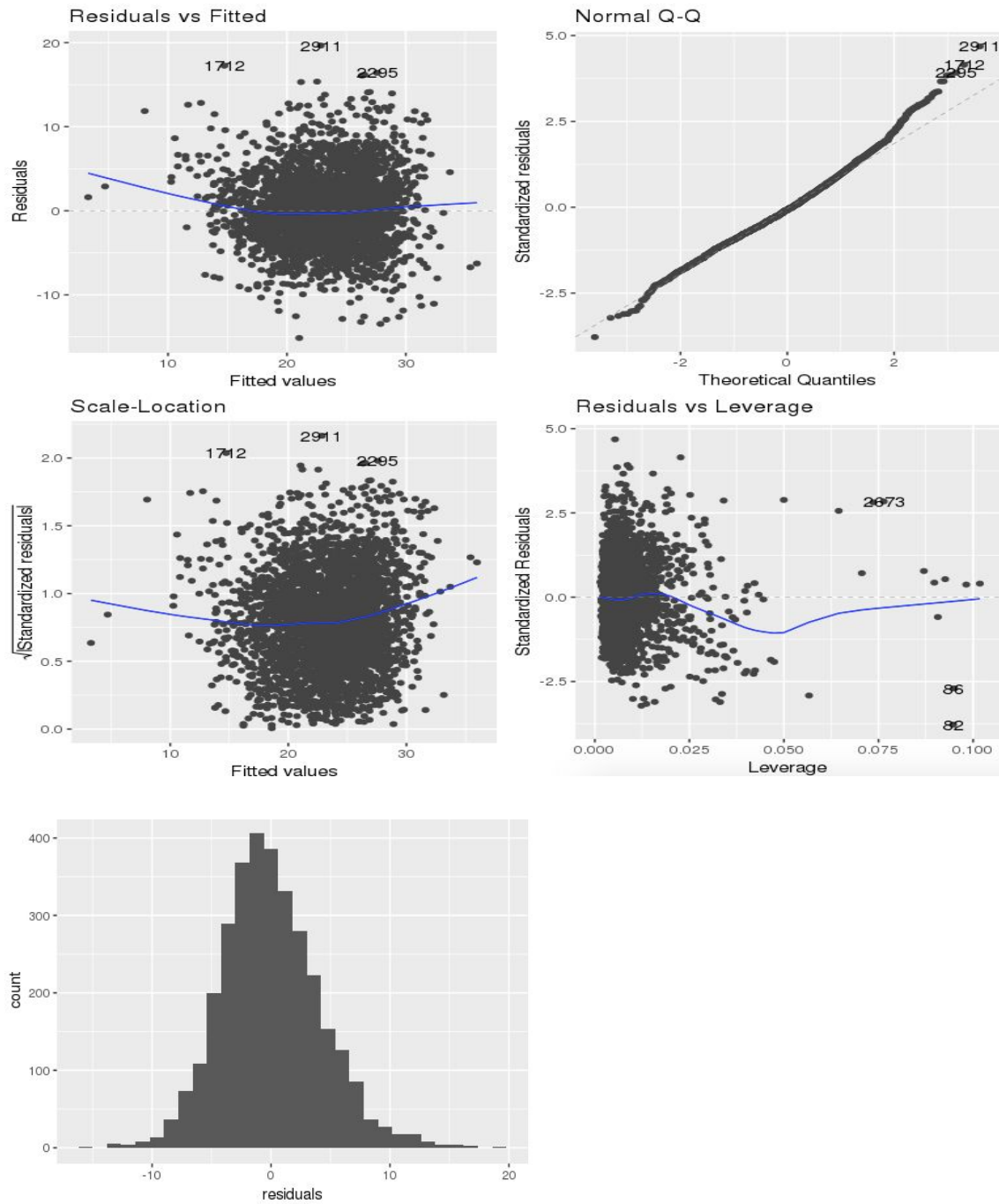

*Figure 11: Regression Tree for Initial Model*

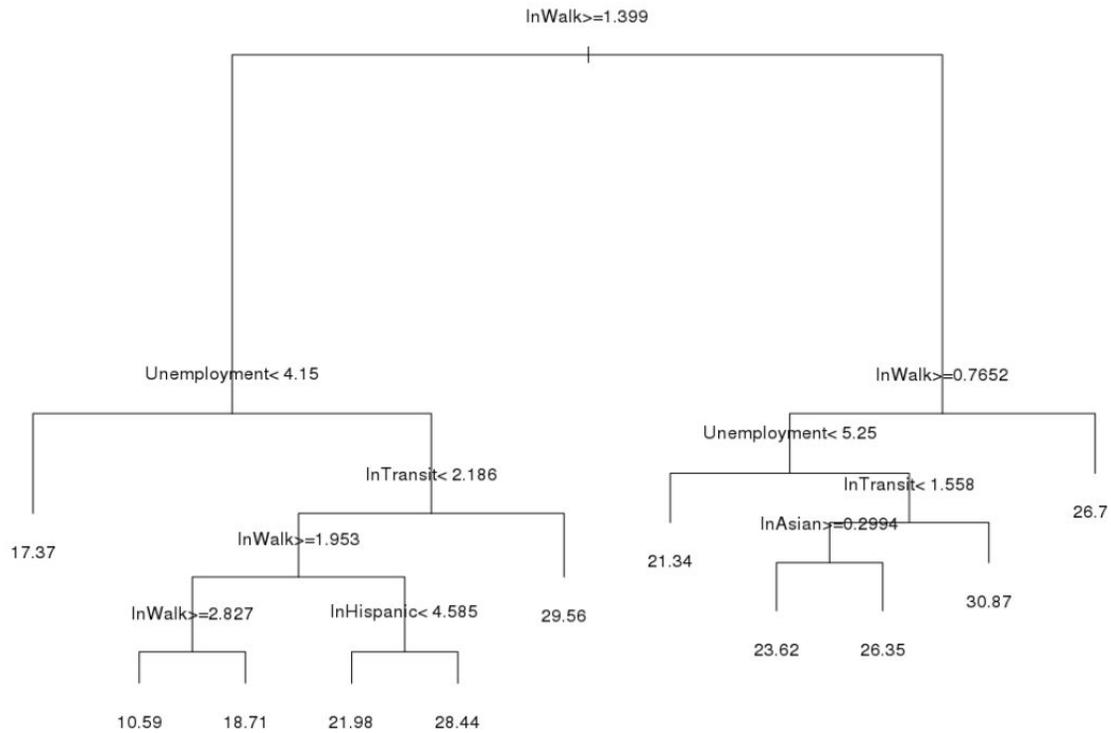

*Figure 12: Multiple Regression Model Summary for Reduced Table after Stepwise (Model 2)*

|    | term | estimate | std.error | statistic | p.value |
|----|------|----------|-----------|-----------|---------|
| 1 | (Intercept) | 2.484179e+01 | 2.228084 | 11.149 | 2.382737e-28 |
| 2 | lnNative | -1.572825e+00 | .118 | -13.323 | 1.874255e-39 |
| 3 | lnTransit | 2.476356e+00 | 2.125535e-01 | 11.650 | 9.438701e-31 |
| 4 | lnPacific | -1.079821e+00 | 5.599260e-01 | -1.928 | 5.388050e-02 |
| 5 | lnTotalPop | 6.863456e-01 | 8.771954e-02 | 7.824 | 6.879078e-15 |
| 6 | lnAsian | -1.233776e+00 | 2.269055e-01 | -5.437 | 5.812315e-08 |
| 7 | lnWalk | -4.495161e+00 | 2.191966e-01 | -20.507 | 6.514366e-88 |
| 8 | lnBlack | -5.168233e-01 | 8.919435e-02 | -5.794 | 7.523961e-09 |
| 9 | lnHispanic | -1.203713e+00 | 9.368381e-02 | -12.848673 | 7.025994e-37 |
| 10 | i.westTRUE | -1.675054e+00 | 2.944419e-01 | -5.688914 | 1.393626e-08 |

| 11 | i.midwestTRUE | -2.171088e+00 | 2.057114e-01 | -10.554048 | 1.275986e-25 |
| 12 | i.otherRegionTRUE | 2.183920e+00 | 5.905067e-01 | 3.698384 | 2.206206e-04 |
| 13 | Unemployment | 3.917735e-01 | 2.854101e-02 | 13.726689 | 1.042592e-41 |
| 14 | Carpool | 5.631487e-02 | 3.204524e-02 | 1.757355 | 7.895284e-02 |
| 15 | Construction | 2.162483e-01 | 2.437852e-02 | 8.870446 | 1.186624e-18 |
| 16 | Poverty | -4.213422e-02 | 1.881474e-02 | -2.239427 | 2.519658e-02 |
| 17 | IncomePerCap | 7.744915e-05 | 2.406034e-05 | 3.218955 | 1.299439e-03 |
| 18 | Drive | -7.858061e-02 | 1.827719e-02 | -4.299382 | 1.763870e-05 |

*Figure 13: Table for reduced MLR model with all interaction terms added (Model 3)*

| Term | Coefficient | Std Error | t-value | p-value |
|---|---|---|---|---|
| (Intercept) | 2.966e+01 | 2.530e+00 | 11.723 | < 2e-16 *** |
| lnNative | -1.250e+00 | 1.199e-01 | -10.425 | < 2e-16 *** |
| lnTransit | 1.376e+00 | 5.869e-01 | 2.345 | 0.019084 * |
| lnPacific | -1.118e+00 | 5.495e-01 | -2.035 | 0.041966 * |
| lnTotalPop | 7.054e-01 | 8.687e-02 | 8.120 | 6.59e-16 *** |
| lnAsian | -1.268e+00 | 4.306e-01 | -2.944 | 0.003263 ** |
| lnWalk | -1.893e+00 | 5.815e-01 | -3.255 | 0.001140 ** |
| lnBlack | -3.384e-01 | 9.001e-02 | -3.760 | 0.000173 *** |
| lnHispanic | -2.022e+00 | 1.576e-01 | -12.831 | < 2e-16 *** |
| i.westTRUE | -1.527e+00 | 2.943e-01 | -5.190 | 2.24e-07 *** |
| i.midwestTRUE | -2.397e+00 | 2.030e-01 | -11.808 | < 2e-16 *** |
| i.otherRegionTRUE | 1.618e+00 | 6.650e-01 | 2.433 | 0.015031 * |
| Unemployment | 3.134e-01 | 6.551e-02 | 4.784 | 1.79e-06 *** |
| Carpool | 3.266e-02 | 3.508e-02 | 0.931 | 0.351957 |
| Construction | 2.225e-01 | 2.391e-02 | 9.304 | < 2e-16 *** |
| Poverty | -8.568e-02 | 1.901e-02 | -4.506 | 6.84e-06 *** |
| IncomePerCap | 2.082e-05 | 2.432e-05 | 0.856 | 0.391964 |
| Drive | -1.142e-01 | 2.379e-02 | -4.800 | 1.66e-06 *** |
| I(lnWalk^2) | -7.328e-01 | 2.039e-01 | -3.594 | 0.000330 *** |
| lnWalk:Unemployment | -4.184e-02 | 3.001e-02 | -1.394 | 0.163371 |
| lnTransit:Unemployment | 6.875e-02 | 3.977e-02 | 1.729 | 0.083952 . |
| lnTransit:lnWalk | -1.093e+00 | 2.549e-01 | -4.287 | 1.86e-05 *** |
| lnHispanic:Unemployment | 8.114e-02 | 1.464e-02 | 5.541 | 3.26e-08 *** |
| lnTransit:lnAsian | 1.800e+00 | 1.830e-01 | 9.832 | < 2e-16 *** |
| lnAsian:Unemployment | -1.518e-01 | 4.552e-02 | -3.335 | 0.000862 *** |
| | | | | |
| Multiple R-Squared: | 0.469 | | Adj R-Sq: | 0.465 |

*Figure 14: Final Multiple Regression Model*

| Term | Coefficient | Std Err | t-value | p-value |
|---|---|---|---|---|
| (Intercept) | 31.66014 | 1.82777 | 17.322 | < 2e-16 *** |
| lnNative | -1.24992 | 0.11981 | -10.433 | < 2e-16 *** |
| lnTransit | 1.46206 | 0.58197 | 2.512 | 0.012045 * |
| lnPacific | -1.11888 | 0.54392 | -2.057 | 0.039761 * |
| lnTotalPop | 0.70983 | 0.08677 | 8.181 | 4.03e-16 *** |
| lnAsian | -1.16539 | 0.42008 | -2.774 | 0.005566 ** |
| lnWalk | -1.75621 | 0.56256 | -3.122 | 0.001813 ** |
| lnBlack | -0.33905 | 0.08996 | -3.769 | 0.000167 *** |
| lnHispanic | -2.01440 | 0.15620 | -12.896 | < 2e-16 *** |
| i.westTRUE | -1.55090 | 0.29338 | -5.286 | 1.33e-07 *** |
| i.midwestTRUE | -2.41798 | 0.20207 | -11.966 | < 2e-16 *** |
| i.otherRegionTRUE | 1.62189 | 0.66248 | 2.448 | 0.014410 * |
| Unemployment | 0.32060 | 0.06489 | 4.941 | 8.18e-07 *** |
| Construction | 0.22350 | 0.02365 | 9.451 | < 2e-16 *** |
| Poverty | -0.09505 | 0.01453 | -6.543 | 7.01e-11 *** |
| Drive | -0.12864 | 0.01849 | -6.958 | 4.17e-12 *** |
| I(lnWalk^2) | -0.80928 | 0.18616 | -4.347 | 1.42e-05 *** |
| lnWalk:Unemployment | -0.04403 | 0.02993 | -1.471 | 0.141405 |
| lnTransit:Unemployment | 0.06284 | 0.03943 | 1.594 | 0.111041 |
| lnTransit:lnWalk | -1.11004 | 0.25390 | -4.372 | 1.27e-05 *** |
| lnHispanic:Unemployment | 0.08089 | 0.01456 | 5.556 | 2.99e-08 *** |
| lnTransit:lnAsian | 1.78222 | 0.17711 | 10.063 | < 2e-16 *** |
| lnAsian:Unemployment | -0.15954 | 0.04488 | -3.555 | 0.000383 *** |
| | | | | |
| Multiple R-Sq | 0.4688 | | Adj R-Sq | 0.4652 |

Poverty is associated with decreased MeanCommute time. Our hypothesis was false. Westerners, midwesterners are associated with lower commute times than Southerners, the baseline variable. Other Regions are associated with higher mean commute time than Southerners.Commute time of Northerners is not significantly different from that of southerners. Higher proportion of number of hispanics is associated with lower commute time.

**Conditions for Final Model:**

Linearity - As visible in the scatterplots of Figure 4 of the appendix, we know that the individual independent variables show approximately linear relationships with mean commute time.

I - In Figure 6, we tested for multicollinearity between initial variables and found very little.

N - As visible in the QQ plot below, the residuals of the model appear to be approximately normal distributed, at least until the very highest of the residual values.

E - Looking at the Residual v Fitted plot below, it appears the spread of the data remained fairly consistent across increasing fitted values.

S - This data comes from the census, which means it fully comprehensive and represents the population.

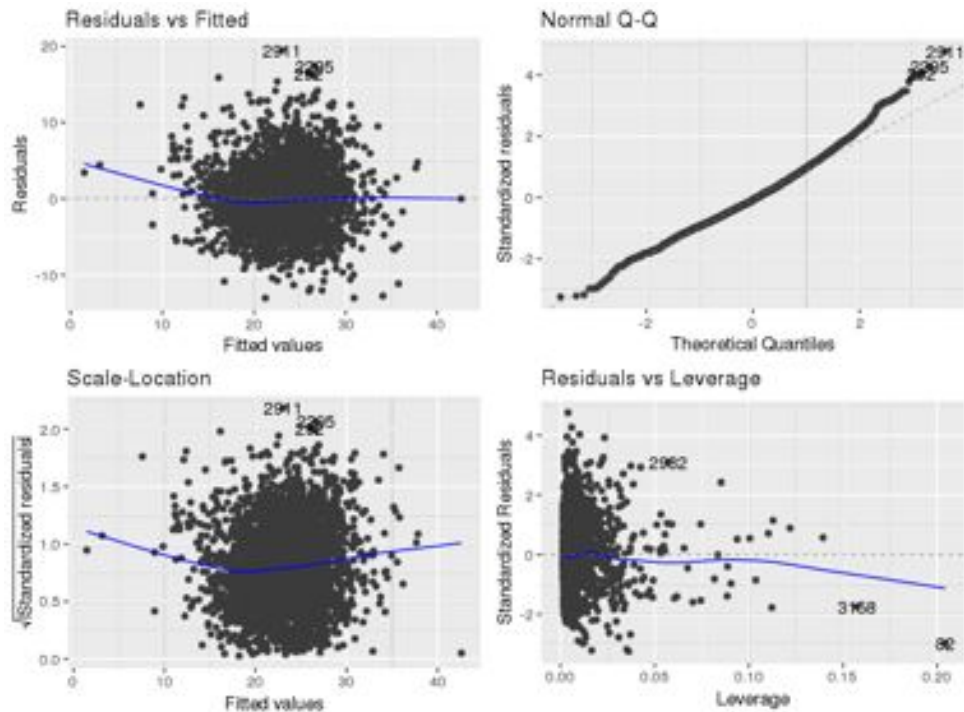

*Figure 16: Comparison of Models*

| Model Name | A.I.C. | R-squared | Adjusted-R-squared | **µ** X-Error(Cross validation error) | # Variables in Model |
|---|---|---|---|---|---|
| Model One:Initial Model | 18415 | 0.4384 | 0.4342 | 17.95708 | 24 |
| Model 2:Step( Initial) | 18405 | 0.4377 | 0.4347 | 17.89115 | 17 |
| Model 3:Step(Initial) Plus All Interaction terms from rtree | 18234 | 0.4690 | 0.4650 | 16.97825 | 24 |
| Final: Stepwise reduction of model 3 | 18232 | 0.4688 | 0.4651 | 16.95901 | 22 |

**Nested ANOVA tests:**

Conclusions of Nested ANOVA comparisons and rankings of models are an equivalence relation that is reflexive, symmetric and transitive. According to ANOVA, Model 1 ≤ Model2 ≤ Model3 ≤ Final Model. The p-values from nested ANOVAs (in order) are .7866, 0, 0.514. The F-statistics in order are 0.5628, 26.88, 0.6653. The first p-value indicates that Model One is as good better than Model 2. The second p-value indicates that Model 3 is as good or better than Model 2. The third p-value indicates that the final model as good or better than Model 3. AIC dropping and Xerror, multiple-r squared increasing mean we made a better model in each step. We made the model based on metrics including AIC and multiple r-squared.

*Figure 17: Cross-Validation Plot*

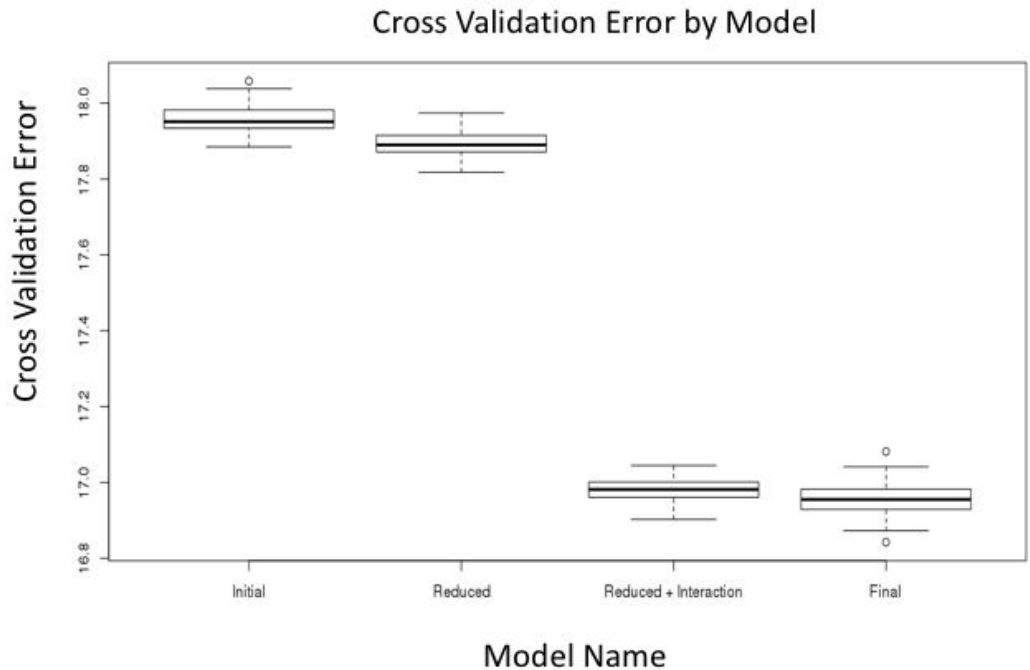

These side-by-side boxplots display distributions for 200 cross validation errors (50 for each model). Their distributions were fairly normal, so their mean values were reported in Table in the Results section.

Multicollinearity does not plague the final model. The first stepwise elimination used to make the second model eliminated multicollinearity. Interaction terms have collinearity with the main effects by default.